\documentclass[aps,prl,twocolumn,showpacs,superscriptaddress]{revtex4-1}
\usepackage{amsmath}
\usepackage{epsfig}
\usepackage{amssymb}
\usepackage{dcolumn}
\usepackage{graphicx}
\usepackage{dcolumn}

\usepackage{color}

\newcommand{\beq}{\begin{equation}}
\newcommand{\eeq}{\end{equation}}

\begin{document}
\date{\today}

\title{Non-conservative Lagrangian method for half-dark solitons in spinor non-equilibrium Polariton condensates}

\author{Florian Pinsker}
\affiliation{Department of Applied Mathematics and Theoretical Physics, University of Cambridge, Wilberforce Road, CB3 0WA,  United Kingdom}

%\affiliation{Clarendon Laboratory, University of Oxford, Parks Road, Oxford OX1 3PU, United Kingdom}

\email{fp278@cam.ac.uk}

\begin{abstract}

 In this work I introduce a powerful analytical method to analyze states of non-equilibrium polariton Bose-Einstein condensates (BEC). It is shown that the condensate wave functions carrying dark solitons and half-dark solitons can be expressed in terms of the natural parameters of the non-equilibrium system, i.e. the growth and decay rates and the interaction parameters. The derivation is based on a non-conservative Lagrangian formalism, which provides ordinary differential equations for the parameters of the dark soliton solutions in their dynamic environment. Explicit expressions for the stationary solution are stated, which significantly differ from those observed in atomic BEC. Subsequently the method is extended to spinor condensates, which yields ordinary differential equations for the parameters of half-dark solitons. Finally a stationary case with explicit expressions for half-dark solitons is presented as well.

\end{abstract}
\pacs{03.65.-w, 05.45.-a, 67.85.Hj, 03.75.Kk}

\maketitle

\emph{Introduction.--} Polariton Bose-Einstein condensates are prime examples for non-equilibrium systems  of coherent quantum matter waves that can be easily controlled experimentally in contrast to atomic BEC \cite{Light,Kasp, Deng,leg}. However, the theoretical description in the mean-field regime is more complicated due to the non-conservative character of the underlying partial differential equations, and so far  effective and simple methods to analyze solutions have been lacking. BEC have been studied for decades particularly due to their intrinsic nonlinearity and relatively simple mathematical description, particularly because these states of coherent quantum matter support excitations such as solitons \cite{Pin6,Pin7,exi,exi2}, quantum vortices, vortex rings or giant vortices \cite{Pin1,Pin2,Pin3,Pin4,Pin5}, which essentially require such nonlinearities \cite{Stri}. It turns out that these excitations depend significantly on the unique features of the underlying particles forming the BEC.

 Polaritons  are quasiparticles formed  within semiconductor microcavities by external illumination via a laser beam, which can be created locally at will either coherently or incoherently \cite{Light, Kasp}. Polaritons manifest a strong coupling between excitons  within the semiconductor created by excitation of the electron into the conduction band and cavity photons driven by the external light source. Excitons are electrons  coupled to oppositely charged holes correspondingly in the semiconductor and are held together by the Coulomb force between them \cite{Deng}. The so called bright states of excitons can interact with light fields \cite{Hopf,Light,Deng} and as the light field oscillates rapidly within the microcavity strong coupling between excitons and the cavity photon can occur and so together they form the so called polariton quasiparticle within the semiconductor microcavity \cite{Weis,Kasp}.  Although electrons and holes are fermions, the composed polaritons obey half-integer spin and therefore can condense to the same ground state macroscopically and form a quasi BEC. As their mass is $10^9$ times lighter than that of rubidium atoms \cite{Kasp}, the critical temperature at which they BE condense is much higher and observation within the Kelvin range have been reported in CdTe/CdMgTe micro-cavities \cite{Kasp,Light} and more recently in flexible polymer-filled microcavities even at room temperatures \cite{Plum}. This makes these states very attractive for applications as the plethora of papers on this matter shows \cite{1,2,3,4,5,6,7,8}.  An interesting feature of polariton condensates is that they can show polarization or incoherence between the two internal quasi-spin states \cite{Light, Pinsker}, which can stem from the polarization of the light source \cite{Pin6,Pinsker}.  This splitting between the two spin components allows the emergence of intrinsically new excitation of the condensate wave function such as half-solitons or half-vortices \cite{Pinsker, Pin6, Hugo}.

One effective theory for the time evolution of the  polariton-condensate wave function in the mean-field regime is the complex Gross-Pitaevskii equation (cGPE) \cite{Wout,Pin6, Dreis}. In contrast to the Gross-Pitaevskii equation for equilibrium BEC \cite{Pita,Gross,Stri}, the cGPE includes complex spatially dependent terms inducing a local gain (for positive sign of the term) or loss (negative sign) of the mass of the condensate wave function. This change in mass for non-equilibrium condensates stems from the relatively short lifetime of polaritions that is according to state-of-the-art experiments between $1-100$ps depending on the quality of the microcavity confining the light field \cite{Deng,ryan, ryan2,Light}. The fleeting existence of polaritons suggests crucial differences for the possible excitations of the condensate, in particular the pattern formation of dark solitons in contrast to those observed in atomic BEC \cite{Pin4, Pin6}. E.g. the gain of mass increases the speed of sound and consequently the first emergence of excitations in superfluid flow  \cite{Pin6}. While analytical solutions describing the dark soliton have been derived within the classic Gross-Pitaevskii theory \cite{tsp, Stri} a rigorous treatment of dark solitons in non-equilibrium polariton condensates will be provided here. Introducing the explicit theory for non-equilibrium dark solitons utilizing a non-conservative Lagrangian formalism and the Kantorovitch method  \cite{method} allows fundamental statements about the amplitude, width, phase and particularly their interplay with pumping and decay parameters. Furthermore I extend this method to spin sensitive polariton condensates and provide the analytical treatment for half-dark solitons, i.e. to half-dark soliton solutions showing strong  pairwise decoherence as predicted in \cite{Pin6}.

%The paper is organized as follows. First I introduce the polariton condensate theory in the mean-field regime. Subsequently I outline the analytical treatment of non-equilibrium dark solitons in terms of a nonconservative Lagrangian method. Finally I discuss the spinor case and summarize my findings. 

\emph{The model.--}  The spinor polariton field $\boldsymbol{\psi}=(\psi_+,\psi_-)$ is governed by cGLEs coupled to a rate equation for the excitonic reservoir $n_R$ \cite{Pin6, Hugo1,Pinsker,Dreis},
\beq
 i \partial_t  \psi_\pm =   \frac{\mathcal H_{\pm}}{\hbar} \psi_\pm
\eeq
\begin{multline}\label{natural}
 \frac{\mathcal H_\pm}{\hbar}  \psi_\pm = \bigg( - \frac{\hbar \nabla^2}{2 m} +  \alpha_1 ( |\psi_\pm|^2 + n_R)  + \\ + \alpha_2 |\psi_\mp|^2  +   \frac{i }{2} \left(\gamma  n_R - \Gamma  \right) \bigg) \psi_\pm.
\end{multline}
$\alpha_1$ denotes the self-interaction strength and $\alpha_2 = -0.1 \alpha_1$ the interaction strength between components. $m$ is the effective mass of the polaritons. $\gamma$ denotes the scattering rate of the reservoir into the condensate and $\Gamma$ the decay rate of condensed polaritons. The reservoir dynamics are
\beq\label{res}
\partial_t n_R   = P-  n_R (\Gamma_R + \gamma ( |\psi_\pm|^2+ |\psi_\mp|^2)).
\eeq
Here $P$ denotes the pumping distribution and $\Gamma_R$ the reservoir decay rate.  For fast reservoir relaxation $\Gamma_R \gg \Gamma$ the reservoir dynamics are much faster than that of the condensate. Therefore the  reservoir \eqref{res}  can be approximated in a first step as \cite{Dreis}
\beq\label{res}
P  \simeq   n_R (\Gamma_R + \gamma ( |\psi_\pm|^2+ |\psi_\mp|^2)).
\eeq

\emph{Single non-equilibrium condensate.--} For coherent spin states the set of equations simplifies to \cite{Dreis,Wout, jona}
\begin{multline}\label{natural2}
 \frac{i}{\hbar} \partial_t  \psi = \bigg( - \frac{\hbar \nabla^2}{2 m}  + \alpha'_1 ( |\psi|^2 + n_R)   + \\  +  \frac{i }{2} \left(\gamma  n_R - \Gamma  \right) \bigg) \psi.
\end{multline}
Here we approximate the reservoir density by
\beq\label{res2}
n_R \simeq \frac{P}{\Gamma_R + \gamma |\psi|^2}.
\eeq
For repulsive condensates the coefficient of the real-valued nonlinear term  is $\alpha'_1 > 0$.
One can further apply the Keeling/Berloff approximation for the complex terms of \eqref{natural2} simplifying the reservoir and decay of polaritons as in \cite{jona, Dreis, Tris}, i.e.
\beq\label{ooo}
  \frac{i }{2} \left(\gamma  n_R - \Gamma  \right) \simeq i ( \delta +  \beta_1 |\psi|^2).
\eeq
I note that in \cite{Polo} the case of bright solitons has been studied for the cGPE with cubic-quintic nonlinearities and I employ and extend this methodology in the following. To proceed we set the sign of the linear complex term $\delta$ to be  positive 
%(stability requires us to introduce a cutoff \cite{jona}
and so $\varepsilon = - |\varepsilon|$ models the density dependent decay. Further we neglect the blueshift from the reservoir. Rescaling \eqref{natural2} $\vec r = \sqrt{2m/\hbar} \cdot \vec x$, $z = \hbar t$ and $\psi \to \frac{1}{\sqrt{\alpha'_1}} E$ and using \eqref{ooo} yields
\begin{equation}\label{natural3}
i \partial_z  E = - \Delta E  +  |E|^2 E + i \delta E + i \varepsilon |E|^2 E.
\end{equation}
with the $1$D Laplacian $\Delta E = \partial_r^2 E$ and $\varepsilon = \beta_1/\alpha_1$. For repulsive self-interactions given that $\delta =0$ and $\varepsilon =0$ eq. \ref{natural3} has a general analytic dark soliton solution \cite{Pin5,parker,tsp,du}, i.e.
\begin{equation}\label{anal}
E(z,r) =  \left( A \cdot  {\rm tanh}(B( r - C z)) + i  D \right) \cdot \exp(i z +K r),
\end{equation}
where $A,B,C,D,K$ are constants.

Following \cite{method,Polo,Polo2} the Lagrangian of our system $\mathbb L$ can be separated into the conservative part $\mathbb L_C$  (real-valued energy terms)  and a dissipative part inducing the non-equilibrium properties of the polariton condensate due to pumping and decay of the polaritons, $\mathbb L_Q$ (complex-valued terms). The total Lagrangian is given by $\mathbb L = \mathbb L_C + \mathbb L_Q$ with the components
\begin{equation}
\mathbb L_C =  \left( \frac{i}{2} \left(E \partial_z E^* - E^* \partial_z E \right) + |\nabla E|^2 + \frac{|E|^4}{2} \right)
\end{equation}
and
\beq
\mathbb L_Q = i  \left( \delta |E|^2+ \varepsilon \frac{|E|^4}{2}  \right).
\eeq
By Hamilton's principle vanishing variation $\delta (\int \int (\mathbb L_C + \mathbb L_Q) dz dr) = 0 $ is obtained for the minimizer, i.e. a solution of the Euler-Lagrange equation $\eqref{natural3}$ that corresponds to  the Lagrangian $\mathbb L$ via \cite{anal}
\beq
\sum_\eta \frac{d}{d \eta} \left( \frac{\partial}{\partial E^*_\eta} \mathbb L - \frac{\partial }{\partial E^*} \mathbb L \right) = 0.
\eeq
We optimize over the transverse coordinate (Kantorovitch method) and get \cite{Polo,Polo2}
\begin{multline}\label{kat}
\int^\infty_{-\infty} dr \left( \sum_\eta \frac{d}{d \eta} \left( \frac{\partial}{\partial E'^*_\eta} \mathbb L - \frac{\partial }{\partial E^*} \mathbb L \right)  \right) = \\ = 2 {\rm Re} \int^\infty_{-\infty} dr \left(  \frac{\partial }{\partial E^*} \mathbb L - \sum_\eta \frac{d}{d \eta} \frac{\partial}{\partial E'^*_\eta} \mathbb L \right)
\end{multline}
with $\eta =r,z$. By denoting $L_c = \int^\infty_{-\infty} dr \mathbb L_c$ the Kantorovitch formula \eqref{kat} can be rewritten as
\begin{multline}\label{into}
\frac{d}{dz} \left(\frac{\partial L_c}{\partial \nu'}  \right) - \frac{\partial L_c}{\partial \nu} = \\ = 2 {\rm Re} \int^\infty_{-\infty}  dr \left(i \delta E + i \varepsilon |E|^2 E \right) \frac{\partial E^*}{\partial \nu}.
\end{multline}
Here $\nu = A(z), W(z), P (z), C(z)$ as we are using a dark soliton ansatz similar to \eqref{anal} for the limiting case $D \to 0$.  To show the non-equilibrium properties let us extend the analytical expression as follows.
%\beq
%E(z,r) = A(z) {\rm sech} \left(\frac{r}{W(z)} \right) \exp(i P (z) + i C(z) r^2).
%\eeq
\begin{multline}\label{Ansatz}
E(z,r) = \\Ê=  A(z) \cdot {\rm tanh} \left(\frac{r}{W(z)} \right) \exp(i P (z) + i C(z) r).
\end{multline}
In order that $L_c$ and the r.h.s. of \eqref{into} is integrable we renormalize the integral by subtracting the constant stationary solution, the procedure for renormalizing the nonlinearity of equilibrium solutions is outlined in \cite{Stri}.  To obtain the non-equilibrium solutions I suggest by analogy to consider the relative pumping strength/decay of the dark soliton solution compared to the constant ground state (GS). Thus
\begin{multline}
\mathbb L^{\rm ren}_C =  \bigg( \frac{i}{2} \left(E \partial_z E^* - E^* \partial_z E \right) + \\ + |\nabla E|^2 + \frac{(|E|^2 - A^2)^2}{2} \bigg)
\end{multline}
\beq\label{two}
\mathbb L^{\rm ren}_Q = i  \left( \delta (|E|^2-A^2) + \varepsilon \frac{(|E|^2 - A^2)^2}{2}  \right).
\eeq
Using \eqref{Ansatz} we obtain the explicit expression for $L^{\rm ren}_c $ (see supplemental material \cite{supp} eq. $1$).
%\begin{equation}
%L^{\rm ren}_c = \frac{2 A^2 (2 + 
%   W^2 (A^2 - 3 (C^2 + P')))}{3 W}.
%\end{equation}
We plug this expression into the l.h.s of eq. \ref{into} and obtain four equations four the four unknowns $\nu$. Variation over $A$ yields employing the abbreviation $' = \partial_z$
% \begin{equation}
%P' = \frac{2 - 8 A^2 W^2 + 
% 4 C^2 \pi^2 W^4 - \pi^2 W^4 C'}{12 W^2}
% \end{equation}
\begin{equation}\label{P}
P' =\frac{2 + 2 A^2 W^2 - 3 C^2 W^2}{3 W^2}
\end{equation}
 and by variation regarding $W$ we obtain
%\beq
%C' = \frac{2 (-1 - 2 A^2 W^2 + 6 C^2 \pi^2 W^4 - 6 W^2 P')}{
% 3 \pi^2 W^4}
% \eeq
\beq\label{W}
W = \pm \frac{\sqrt{2}}{\sqrt{A^2 - 3 C^2 - 3 P'}}.
 \eeq
%Then variation with regards to $C$ we obtain
%\begin{multline}
%\frac{-\pi^2}{12}  \partial_z (A^2 W^3) - \frac{2}{3} A^2 C \pi^2 W^3 = \\ = \frac{1}{18} A^2 (3 \delta \pi^2 + 2 A^2 \varepsilon (-6 + \pi^2)) W^3
%\end{multline}
Variation regarding $P$ yields
\beq
(-A^2 W)' = -2 A^2 \delta W + \frac{4}{3} A^4 \varepsilon W,
\eeq
and variation over $C$ gives
\begin{equation}\label{C}
-4 A^2 C W =0.
\end{equation}
To satisfy \eqref{C} we set $C=0$ and so we effectively consider a set of three equations for the three unknows $A,W,P$. For vanishing time dependence of $W$ and $A$, $(-A^2 W)'  =0$. Thus one obtains a simple formula for the amplitude of the dark soliton in the non-equilibrium condensate in terms of linear pumping $\delta$ and density dependent decay $\varepsilon = - |\varepsilon|$ parameters,
\beq\label{p}
A = \pm \frac{\sqrt{ 3 \delta}}{\sqrt{2 \varepsilon}}.
\eeq
This confirms the intuitive behavior of the polariton condensate: As the lifetime of polaritons (for different realizations and e.g. microcavities) decreases the amplitude of the stationary state decreases and as the pumping strength increases the amplitude  of the excited state does.
In addition for the width of the soliton we get using \eqref{P}, \eqref{W} and \eqref{p},
\beq
W = \pm  i\frac{2}{\sqrt{A^2}} = \pm i \sqrt{\frac{8}{3} \frac{\varepsilon}{\delta}}.
\eeq
This formula shows that the higher the pumping strength $\delta$, the smaller the width of the dark soliton and stronger decay implies a wider soliton.
Using the formulas for the width and the amplitude we derive the time dependent phase $P$ given through \eqref{P}, i.e.
\beq
P(z) = \frac{2 (1 + A^2 W^2) z}{3 W^2} + const. = \frac{3}{4} \frac{ \delta}{ \varepsilon}  z + const.
\eeq

\emph{Non-equilibrium spinor condensate.--}  Let us now consider the full spinor system \eqref{natural} for one spatial dimension neglecting the secondary effects of TMTE splitting and energy relaxation  \cite{Pin6}.
%The r.h.s. is given by
%\begin{multline}
% \frac{\mathcal H_\pm}{\hbar}  \psi_\pm = \bigg( - \frac{\hbar \nabla^2}{2 m} + \alpha_1 ( |\psi_\pm|^2 + n_R) + \alpha_2 |\psi_\mp|^2  + \\ + U +  \frac{i }{2} \left(\gamma  n_R - \Gamma  \right) \bigg) \psi_\pm +  \frac{H_x}{2} \psi_\mp
%\end{multline}
As in the previous derivation we use the Keeling/Berloff approximation \eqref{ooo} and generalize this expression to the spinor condensate case by
\beq\label{ooo2}
  \frac{i }{2} \left(\gamma  n_R - \Gamma  \right) \simeq i ( \delta + \varepsilon \alpha'_1 (|\psi_+|^2+|\psi_-|^2)).
\eeq
Applying a similar rescaling as for  \eqref{natural3}, we obtain
\begin{multline}\label{natural4}
i \partial_z  E_\pm = - \Delta E_\pm  +  (|E_\pm|^2 + \gamma  |E_\mp|^2) E_\pm  \\+ i \delta E_\pm + i \varepsilon (|E_\pm|^2 + \gamma_2 |E_\mp|^2) E_\pm.
\end{multline}
Here $\gamma = \alpha_2/\alpha_1$ is the magnitude of the elastic cross-scattering between the two spin components and $\gamma_2$ models misalignment between self- and cross-spin decay. The associated non-renormalized Lagrangians (for dissipative and conservative part) for the spinor condensate are
\begin{multline}
\mathbb L^{\rm ren}_C = \sum_{k \in \{+, - \}}  \bigg( \frac{i}{2} \left(E_k \partial_z E^*_k - E^*_\pm \partial_z E_k \right) + |\nabla E_k|^2  +\\ + \frac{|E_k|^4}{2}  \bigg) +  \gamma  |E_{+}|^2 |E_-|^2,
\end{multline}
 including  cross-interaction terms in contrast to the coherent states formalism from above, and
%\begin{multline}
%\mathbb L^{\rm ren}_Q =\sum_{k \in \{+, - \}} i \bigg( \delta (|E_k|^2 -A_k^2)+  \varepsilon \frac{(|E_k|^2-A^2_k)^2}{2}  \bigg) + \\  +  i \varepsilon \gamma_2  (|E_{+}|^2 -A^2_{+}) (|E_-|^2 -A_-^2),
%\end{multline}
\begin{multline}\label{last}
\mathbb L^{\rm ren}_Q =\sum_{k \in \{+, - \}} i \bigg( \delta (|E_k|^2 -A_k^2)+  \varepsilon \frac{|E_k|^4}{2}  \bigg) + \\  +  i \varepsilon \gamma_2  |E_{+}|^2 |E_-|^2,
\end{multline}
which via the last term in \eqref{last} takes into account the effects of cross-decay or interspecific competition between the two spin components.
Let us now adapt ansatz \eqref{Ansatz} to the spinor case and write $\Psi' = (E_+,E_-)$ with
\begin{multline}
E_\pm(z,t) =  \\ =   A_\pm(z) \cdot {\rm tanh} \left(\frac{r \pm r_0}{W_\pm(z)} \right) \exp(i P_\pm (z) + i C_\pm (z) r).
\end{multline}
%and
%\begin{multline}
%E_-(z,t) = \\ = A_-(z) \cdot {\rm tanh} \left(\frac{r-r_0}{W_-(z)} \right) \exp(i P_- (z) + i C_- (z) r).
%\end{multline}
In this ansatz component $+$ is assumed to carry a dark soliton with density minimum at $r=-r_0$ while  component $-$ carries a dark soliton with a minimum at $r=r_0$. The form of the ansatz resembles the half-dark soliton splitting observed e.g. in \cite{Hu2, Pin6}. I further assume both solitons to be far away from each other, i.e. $r_0 \gg 0$, to obtain converging integrals particularly for the energy density of the cross-interaction term $\gamma  |E_\mp|^2 |E_\pm|^2$. Thus we can approximate the cross-component effects via a dark soliton solution in one component and a spatially homogeneous solution $A_\pm$ in the other component, which is justified by the relation $\tanh^2 (r) \simeq 1$ when $r \gg 0$. Note that for the spinor case we again write $L_c = \int^\infty_{-\infty} dr \mathbb L_c$ and $L_q = \int^\infty_{-\infty} dr \mathbb L_Q$.  Hence, the integral of the conservative Lagrangian is formally
\begin{multline}
L'^{\rm ren}_c =  \int^\infty_{-\infty} dr \sum_{k \in \{+, - \}}  \bigg( \frac{i}{2} \left(E_k \partial_z E^*_k - E^*_\pm \partial_z E_k \right) + \\ + |\nabla E_k|^2  + \frac{(|E_k|^2-A^2_k)^2}{2}   \bigg) + \gamma C
\end{multline}
with the approximation, which includes the renormalization introduced in \cite{Stri}, i.e.
\begin{multline}
C \simeq  \int^\infty_{0} dr (|E_-|^2 - A^2_-) |A_+|^2 + \\ +
\int^0_{-\infty} dr (|E_+|^2 - A^2_+) |A_-|^2.
\end{multline}
In analogy the non-conservative part is
\begin{multline}
L'^{\rm ren}_q = \int^\infty_{-\infty} dr \sum_{k \in \{+, - \}} i \bigg( \delta (|E_k|^2 -A_k^2)+ \\ + \varepsilon \frac{(|E_k|^2-A^2_k)^2}{2}   \bigg) + i \varepsilon \gamma_2  C,
\end{multline}
which takes an additional cross-decay term $\varepsilon \gamma_2  C$ into account.
The result for the integrated Lagrangian for the half-dark soliton spinor solution is
\begin{widetext}
\begin{equation}\label{spin}
L'^{\rm ren}_c = \frac{2 A^2_+ (2 + 
   W^2_+ (A^2_+ - 3 (C^2_+ + P'_+)))}{3 W_+} +  \frac{2 A^2_- (2 + 
   W^2_- (A^2_- - 3 (C^2_- + P'_-)))}{3 W_-}  - \gamma A^2_+ A^2_-  (W_+ + W_-).
\end{equation}
\end{widetext}
For the generalized spinor condensate the parameters of variation are $\nu = A_\pm (z), W_\pm (z), P_\pm (z)$ and $C_\pm (z)$.
By variation regarding $A_\pm$ and $W_\pm$ we obtain the corresponding parameter equations
\begin{equation}\label{P2}
P'_\pm =\frac{2 + 2 A^2_\pm W^2_\pm - 3 C^2_\pm W^2_\pm - \frac{1}{2} \gamma (A_\mp^2 W^2_\pm + A_\mp^2 W_\pm W_\mp)}{3 W^2_\pm}.
\end{equation}
and
\beq\label{W2}
W_\pm = \pm\frac{2}{\sqrt{2 A^2_\pm - 6 C^2_\pm - 3 A_\mp^2 \gamma - 6 P'}}
\eeq
Variation regarding $P$ yields
\begin{multline}\label{A2}
(-A^2_\pm W_\pm)' =  -2 A^2_\pm \delta W_\pm + \frac{4}{3} A^4_\pm \varepsilon W_\pm + \\ + \gamma_2 \varepsilon 2 A^2_\pm A^2_\mp  (W_\pm + W_\mp)
\end{multline}
while variation over $C$ implies
\begin{equation}\label{C2}
-4 A^2_\pm C_\pm W_\pm = 2 \varepsilon \gamma_2 A_\pm^2 A_\mp^2 \log(2)(W_\pm^2 - W_\mp^2).
\end{equation}
As $\gamma \to 0$ and $\gamma_2 \to 0$ we restore the coherent case discussed above.
%For stationary density profiles eq. \ref{A} yields
%\beq\label{A2}
%A_\pm = \pm \left( \frac{\sqrt{ 3 \delta }}{\sqrt{2 \varepsilon}} + \varepsilon \gamma_2 \frac{\sqrt{3 A_\mp^2 (W_\pm +W_\mp) }}{2 \sqrt{\varepsilon W_\pm}}  \right).
%\eeq
%Let us further 
For stationary density profiles and assuming the width of both half-dark solitons to be equal  \eqref{C2} implies that $C_\pm = 0$ as long as $A_\pm \neq 0$ and $W_\pm \neq 0$. Then by using eq. \ref{A2}  we obtain the simple solution for the coupled amplitudes of half-dark solitons,
\begin{equation}\label{A22}
A_\pm =  \pm \sqrt{ \frac{ 3 \delta  }{\varepsilon (2- \varepsilon \gamma_2) }}.  
\end{equation}
%\begin{equation}\label{A22}
%A_\pm = \pm \left( \frac{\sqrt{ 3 \delta }}{\sqrt{2 \varepsilon}} + \varepsilon \gamma_2 \frac{\sqrt{6 A_\mp^2  }}{2 \sqrt{\varepsilon}}  \right) =  \pm \sqrt{ \frac{ 3 \delta  }{\varepsilon (2- \varepsilon \gamma_2) }}.  
%\end{equation}
Eq. \ref{A22} shows that the cross-decay reduces the amplitude of each component.  The corresponding expressions for the width and the phase of the non-equilibrium half-dark solitons are provided within the supplemental material \cite{supp}. Higher interspecific competition $\gamma_2$  or lowered cross-interactions $\gamma$ widen the half-dark solitons within each spin component of the polariton BEC.

\emph{Conclusions.--} In this work I have introduced a non-conservative  Lagrangian treatment of non-equilibrium polariton condensates within semiconductor microcavities. I have discussed the case of dark solitons and half-dark solitons and have derived ordinary parameter differential equations describing their dynamical behavior. Assuming the system to obey a stationary density profile explicit two parameter expressions for the amplitude, width and phase of the non-equilibrium dark soliton could be stated. These expressions only dependent on the pumping strength $\delta$ and decay parameter $\varepsilon$ of the non-equilibrium condensate. This significantly contrasts the notions of dark solitons in atomic BEC, where the amplitude, phase and width relate to the self-interaction strength, the chemical potential and the mass of the atoms \cite{Stri}. Furthermore the case of non-equilibrium spinor condensates has been considered for the simple scenario of two spatially separated  half-dark soliton as excitations of the corresponding spin components. Again explicit parameter equations could be derived by means of a novel generalized non-conservative Lagrangian method. In the course of this work a renormalization procedure for the non-conservative terms has been introduced in analogy to the well-established formalism for energy renormalization.

\emph{Acknowledgements.--}
I have been financially supported through an EPSRC doctoral prize fellowship at the University of Cambridge and
%, a David Crighton Fellowship held at the University of Oxford  and by the King Abdullah University of Science and Technology (KAUST) Award No. KUK-I1-007-43. 
I would like to thank Hugo Flayac for stimulating discussions.

 \end{document}